\documentclass[twocolumn]{article}
\usepackage{amsmath, amssymb, latexsym, epsfig}
\begin{document}
\newcommand{\di}{\displaystyle}
\newcommand{\reff}[1]{(\ref{eq:#1})}
\newcommand{\labl}[1]{\label{eq:#1}}
\newcommand{\Grad}{\operatorname{grad}}
\newcommand{\Div}{\operatorname{div}}
\newcommand{\Rot}{\operatorname{rot}}
\renewcommand{\thefootnote}{\fnsymbol{footnote}}
 
\noindent{\Large\bf The Rotating Magnet}

\vspace{2mm}

\noindent P. Hrask\'{o}\footnote{peter@hrasko.com}
 
\vspace{5mm}

{\bf Abstract:} Axisymmetric permanent magnets become electrically
polarized due to their rotation around the symmetry axis. This
phenomenon is considered in detail for both conducting and dielectric magnets.
The results are applied to the Earth which is predicted to be electrically polarized. 
It is suggested that this polarization can be detected in a tethered satellite 
experiment.

\vspace{5mm}

\renewcommand{\thefootnote}{1} 
\noindent{\large\bf The World Dynamo}\hfill
 
\vspace{5mm}

Many years ago I read as a schoolboy an exciting book about a dogged
ingeneer who took it into his head to lay down a massive electric cable
along a large segment of a meridian, for, according to his calculations,
the rotation of the Earth in its own magnetic field should induce in the
cable
currents of enormous strength and this "world dynamo" --- that was the
name of the book --- would supply mankind with cheep electricity.
 
Though I was charmed with the idea I had something on my mind: What if
the magnetic field rotated together with the Earth? If it did the cable
would never cross the lines of force of the field and no current would
be induced. I did not realized until much later
that my question itself was rather problematic since
the meaning of rotation of an axisymmetric magnetic field around its
symmetry axis was far from being obvious. If lines of force existed
in reality and the  motion of either of them could be followed in time, my
question would be all right and could in principle be answered. But
these lines are only mathematical abstractions deviced to aid the
visualization of the field structure and have no real existence. The
magnetic field of a magnet is in fact independent of whether the magnet
rotates around its symmetry axis or not --- in this respect the engineer
was certainly right.

In what follows we will consider a homogeneous spherical magnet of
radius $a$, rotating with a constant angular velocity
$\boldsymbol\Omega$, whose magnetization density $\boldsymbol M$ is
parallel to the axis of rotation (assumed to be the
$z$-axis). The cable of the world dynamo will be represented by a rigid
linear conductor $\cal L$, not necessarily a plane curve, which
connects the "north pole" and the
"equator" of the sphere. The conductor $\cal L$ 
may also rotate with an angular velocity $\boldsymbol\omega$ which is
parallel to $\boldsymbol\Omega$ but may differ from it in magnitude.
(In the world dynamo we have $\boldsymbol\omega = \boldsymbol\Omega$ and
$\cal L$ lies along a meridian
on the surface of the Earth but it will turn out expedient to
deal with the more general case.) 
At the endpoints $A$ 
($\vartheta = 0^\circ$) and $B$ ($\vartheta = 90^\circ$) the
linear conductor is connected electrically
to the magnet by means of sliding
contacts so as to make $\cal L$ part of an electric
circuit closed through the magnet.

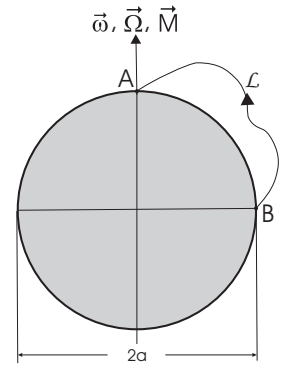
\vspace{2mm}

%\psdraft
\begin{center}
\epsfig{file=fig.eps,width=35mm}
\end{center}
%\psfull

\vspace{2mm}

Below we will restrict ourselves to the discussion of the physical basis,
underlying the world dynamo idea. It will be left to the reader
to judge whether such an extraordinary power
plant if realized in practice would indeed
be continuously supplying electric power or not.

\vspace{5mm}

\noindent{\large\bf The Rotating Conducting Magnet}\hfill

\vspace{3mm}

Consider a rotating metallic magnet of conductivity $\gamma$
temporarily stripped of
the linear conductor $\cal L$. In a conductor at rest the
connection between the current density $\boldsymbol J$ and the electric
field $\boldsymbol E$ is given by the Ohm's law $\boldsymbol J =
\gamma\boldsymbol E$. When the conductor is moving the electric field
must be supplemented by the electromotive force
$(\boldsymbol V\times\boldsymbol B)$ and in this more general case the
Ohm's law becomes
\begin{equation}
\boldsymbol J = \gamma [\boldsymbol E + (\boldsymbol V\times\boldsymbol B)].\labl{B1a}
\end{equation}
Let us choose the origin of the coordinate system at the center of the
sphere. Then the element of the magnet at $\boldsymbol r$ will have the
velocity
\begin{equation}
\boldsymbol V = (\boldsymbol\Omega\times\boldsymbol r).\labl{B1b}
\end{equation}

From this formula it is obvious that we are working in the {\em inertial
system} in which the center of the magnet is at rest rather than in the
system, rotating together with the magnet around this point. In what
follows we will never replace our reference frame with the corotating
one.

Just as it is in the case of a conductor at rest the current density in a
rotating conducting sphere also vanishes under stationary conditions. In
the latter case, however, the electric field does not disappear together
with the current density since when $\boldsymbol J = 0$ we obtain
from \reff{B1a} the electric field
\begin{equation}
\boldsymbol E = -(\boldsymbol V\times\boldsymbol B)\qquad (r<a)\labl{B2a}
\end{equation}
which is associated with some definite volume and surface charge densities $\rho$
and $\sigma$. Eq. \reff{B2a} and the Maxwell-equation
$\Div\boldsymbol E = \rho /\epsilon_0$ determine the volume charge
density:
\[\rho = -\epsilon_0\Div (\boldsymbol V\times\boldsymbol B).\]

As it is known from magnetostatics the induction within a homogeneously
magnetized sphere is equal to
\begin{equation}
\boldsymbol B = \frac{2}{3}\mu_0\boldsymbol M\qquad (r<a).\labl{B3b}
\end{equation}
Therefore,
\[(\boldsymbol V\times\boldsymbol B) = \frac{2}{3}\mu_0(\boldsymbol V\times\boldsymbol M)\]
and, using \reff{B1b}, we obtain
\begin{equation}
\begin{aligned}
(\boldsymbol V\times\boldsymbol M) &= \bigl ((\boldsymbol\Omega\times\boldsymbol r)\times\boldsymbol
M\bigr ) = \\
& = (\boldsymbol M\cdot\boldsymbol\Omega )\boldsymbol r - (\boldsymbol M\cdot\boldsymbol
r)\boldsymbol\Omega .
\end{aligned}\labl{B3d}
\end{equation}
Let us take now into account that the constant vectors
$\boldsymbol M$ and $\boldsymbol\Omega$ are parallel to each other, 
$\Div\boldsymbol r = 3$ and, finally, 
\[\Div\bigl ((\boldsymbol M\cdot\boldsymbol
r)\boldsymbol\Omega\bigr ) = M\Omega .\] 
Then
\begin{gather}
\Div (\boldsymbol V\times\boldsymbol M) = 2M\Omega\labl{B3c}
\intertext{and}
\rho = -\frac{4}{3}\epsilon_0\mu_0M\Omega .\labl{B3a}
\end{gather}
Using \reff{B3b} it is easy to show that for the electric field \reff{B2a}
$\Rot\boldsymbol E = 0$ which is the second Maxwell-equation for
$\boldsymbol E$ when the fields are constant in time.

The surface charge density can be calculated as in electrostatics. We
have
\begin{equation}
\sigma = \epsilon_0(E_r^+ - E_r^-),\qquad (r=a)\labl{B4a}
\end{equation}
where $E_r^+$ and $E_r^-$ are the radial components of the electric
field on the outer (+) and inner (--) side of the surface of the magnet.
Our previous formulae permit us to write
\begin{equation}
E_r^- = -(\boldsymbol V\times\boldsymbol B)_r = -V_\vartheta B_\varphi + V_\varphi
B_\vartheta .\labl{B4c}
\end{equation}
Since $\boldsymbol B$ has only a $z$-component we have
\begin{equation}
B_\varphi = 0\qquad\text{and}\qquad B_\vartheta =
-\frac{2}{3}\mu_0M\sin\vartheta .\labl{B4b}
\end{equation}
The only nonzero component of
$\boldsymbol V$ is
$V_\varphi$ which is equal to $\Omega r\sin\vartheta$. Hence
\begin{equation}
E_r^- = -\frac{2}{3}a\mu_0M\Omega\sin^2\vartheta\qquad (r=a).\labl{B5a}
\end{equation}
In order to calculate $E_r^+$ the electrostatic potential $\Phi$ outside
the sphere must be known. From the potential the electric field is
obtained as a gradient:
\begin{equation}
\boldsymbol E = -\boldsymbol\nabla\Phi .\labl{B5b}
\end{equation}
Outside the sphere the charge density is zero and $\Phi$ obeys the
Laplace-equation
\begin{equation}
\triangle\Phi = 0,\qquad (r>a)\labl{B5c}
\end{equation}
from the solution of which $E_r^+$ can be calculated as
\begin{equation}
E_r^+ = -\left.\frac{\partial\Phi}{\partial r}\right |_{r=a}.\labl{B5f}
\end{equation}

In Appendix 1 we show that
\begin{gather}
\Phi = -\frac{1}{9}a^5\mu_0M\Omega\cdot\frac{1}{r^3}(3\cos^2\vartheta -
1)\qquad (r\ge a),\labl{B5d}
\intertext{from which we obtain}
E_r^+ = -\frac{1}{3}a\mu_0M\Omega (3\cos^2\vartheta - 1).\labl{B5e}
\end{gather}
Therefore, the surface charge density is given by the
formula
\begin{equation}
\sigma = \frac{1}{3}a\epsilon_0\mu_0M\Omega (1 + \cos^2\vartheta
).\labl{B6a}
\end{equation}
It is straightforward to show that the total surface charge compensates
exactly the total volume charge.

Since both the induced volume and surface charges rotate together with the
magnet they give rise to corresponding current densities. In the case of
the volume charge density this current density is equal to
\begin{equation}
\Delta\boldsymbol J = \rho\boldsymbol V = -\frac{4}{3}\epsilon_0\mu_0M\Omega\boldsymbol
V.\labl{B7a}
\end{equation}
This $\Delta\boldsymbol J$ does not contribute to the l.h.s. of the
Ohm's law \reff{B1a} because it arises from the rotation rather from the
effect of the field $\boldsymbol E$. However, 
$\Delta\boldsymbol J$ generates,
through the Maxwell-equation
$\Rot\Delta\boldsymbol B = \mu_0\Delta\boldsymbol J$, the magnetic field
$\Delta\boldsymbol B$ which must be added to $\boldsymbol B$ on the
r.h.s. of \reff{B1a}. As a consequence this same correction appears on
the r.h.s. of \reff{B2a} also but, as we now show, gives only
vanishingly small contribution to $\boldsymbol E$.

Dimensional considerations based on both physical and geometrical
dimensions lead to the solution
$\vert\Delta\boldsymbol B\vert\sim
a\mu_0\vert\Delta\boldsymbol J\vert$ ($\sim$ denotes "order of magnitude
equality"). Hence, taking into account \reff{B7a} and \reff{B3b}, we
have
\[\left\vert\frac{\Delta\boldsymbol B}{\boldsymbol B}\right\vert\sim
a\mu_0\frac{\epsilon_0\mu_0M\Omega V}{\mu_0M} = \frac{a\Omega V}{c^2}\sim 
\frac{V^2}{c^2}.\]
This correction is indeed very small for both laboratory magnets and 
celestial bodies and so will be neglected. The factors
$(1 - V^2/c^2)^{1/2}$ which should be included at certain places into
our formulae are left out of consideration for the same reason.

\newpage

\vspace{5mm}

\noindent{\large\bf The Rigidly Fixed Contour}\hfill

\vspace{3mm}

Let us assume now that the contour $\cal L$ is rigidly fixed to the
magnet ($\boldsymbol\omega = \boldsymbol\Omega$). If it did not rotate
the electromotive force in it would be given by the formula
\[{\cal E} = \int_{\cal L}\:\boldsymbol E\cdot d\boldsymbol l.\]
For a rotating contour an electromotive force induced by the motion also
contributes to $\cal E$:
\begin{equation}
{\cal E} = \int_{\cal L}\bigl [\boldsymbol E + (\boldsymbol V\times\boldsymbol B)\bigr ]\cdot
d\boldsymbol l.\labl{C1a}
\end{equation}

When calculating the integral the contour must be assumed fixed in our
coordinate system since its motion is already taken into account by the
second term of the integrand\footnote{This method of calculation is
justified if the displacement of the contour is negligible during the
time interval the electromagnetic signal passes through it. Problems
which we are interested in do not require higher accuracy.}.

In \reff{C1a} the electromotive force is a sum of an electric component
$\di{\cal E}_e = \int_{\cal L}\:\boldsymbol E\cdot d\boldsymbol l$ 
and a magnetic (or motional) component $\di{\cal E}_m
= \int_{\cal L}\:(\boldsymbol V\times\boldsymbol B)\cdot d\boldsymbol
l$. Since the former is originated from the electric polarization of the
rotating magnet it is equal to the potential difference
\begin{equation}
{\cal E}_e = \Phi_B - \Phi_A,\labl{C2d}
\end{equation}
which, for fixed endpoints, is independent of the form of the contour.

But ${\cal E}_m$ is contour independent as well since for a closed contour
\begin{equation}
\oint (\boldsymbol V\times\boldsymbol B)\cdot d\boldsymbol l = 0.\labl{C2a}
\end{equation}

This is the consequence of the Stokes-theorem
\[\oint (\boldsymbol V\times\boldsymbol B)\cdot d\boldsymbol l = \int_{\Sigma}\:\Rot (\boldsymbol
V\times\boldsymbol B)\cdot\boldsymbol n\:d\Sigma,\]
in which $\Sigma$ is any surface bounded by the closed contour and
$\boldsymbol n$ is its normal vector. In Appendix 2 we will prove that at any
point of space (i.e. both inside and outside the sphere) the equality
\begin{equation}
\Rot (\boldsymbol V\times\boldsymbol B) = 0\labl{C2b}
\end{equation}
holds from which \reff{C2a} and the contour independence of
${\cal E}_m$ follow (for fixed endpoints).

Since both ${\cal E}_e$ and ${\cal E}_m$ are contour independent the
same is true for the total electromotive force $\cal E$ too, therefore, the
contour in \reff{C1a} may be chosen for convenience. The best choice is
to direct it entirely within the magnet since, according to \reff{B2a},
along such a contour the integrand of \reff{C1a} vanishes. Hence we
conclude that in any linear contour which rotates together with the
magnet and whose endpoints lie on the surface no electromotive force is
induced. As it follows from  Appendix 2 this conclusion remains valid also for an
axisymmetric conducting magnet of any form, rotating around its symmetry
axis.

\vspace{5mm}

\noindent{\large\bf The Unipolar Induction}\hfill

\vspace{3mm}

When $\cal L$ rotates with respect to the magnet
$(\boldsymbol\omega\not=\boldsymbol\Omega )$
the electromotive force in it consists of the same kind of terms as in
the corotating contour. For ${\cal E}_e$ \reff{B5d} remains valid. The
potentials $\Phi_A$ and $\Phi_B$ can be calculated from
\reff{B5d}. In $A$ and $B$ we have $\vartheta =0^\circ$ and
$90^\circ$ respectively and in both cases $r=a$, therefore
\begin{equation}
{\cal E}_e = \Phi_B - \Phi_A = \frac{1}{3}a^2\mu_0M\Omega .\labl{D1a}
\end{equation}
As we saw in the preceding section for a corotating contour
${\cal E}_m = -{\cal E}_e$. This electromotive force depends on the
velocity of the points of the contour in the reference frame chosen and
in the case of corotation it is proportional to $\Omega$. Then, for a
contour, rotating independently of the magnet, ${\cal E}_m$ coincides
with the {\em negative} of \reff{D1a} in which $\Omega$ is replaced by
$\omega$:
\begin{equation}
{\cal E}_m = -\frac{1}{3}a^2\mu_0M\omega .\labl{D1b}
\end{equation}
Therefore, the full electromotive force is given by the equation
\begin{equation}
{\cal E} = \frac{1}{3}a^2\mu_0M(\Omega - \omega ).\labl{D1c}
\end{equation}
Owing to the sliding contacts, $\cal L$ closes through the magnet and
since the latter's conductivity is different from zero a current will
flow in $\cal L$.

According to \reff{D1c} electromotive force and current arise even in a
contour at rest ($\boldsymbol\omega = 0$). This phenomenon known as the
{\em unipolar induction} is rather paradoxical since the current can
explain neither by the law of induction (since the magnetic field is
constant in time) nor as a motional induction (since the contour is at
rest). Unipolar induction originates solely from the electric
polarization of the magnet. For a spherical magnet its magnitude can be
calculated from \reff{D1a} but the sphericity is, of course, not
essential for the phenomenon to occur. In a Faraday-disk
magnetized along its axis of rotation current will be generated even
in the absence of an external magnetic field.

As it is seen from \reff{D1c} the electromotive force depends on the
{\em relative} rotation of the magnet and the contour. This is quite an
unexpected result since rotation is {\em
absolute}: A deformable sphere, rotating in an inertial frame takes on
the shape of an ellipsoid of rotation. This deformation is the
manifestation of the absolute rotation since it exists irrespective 
of the frame of reference from which the sphere is observed.

The role of rotation in electrodynamics is by no means different. The electric
polarization of the rotating magnet is an objective (absolute)
phenomenon in the same sense as the deformation of a rotating sphere
since it demonstrates unequivocally that it is the magnet --- and not
the contour --- which is rotating. Curiously enough, {\em in the special
case of the electromotive force in} $\cal L$ it is only the relative
rotation which counts. But this is so only when the magnet conducts
electricity. For a magnet made of insulator relation \reff{D1c} ceases to
be
valid and $\cal E$ turns out to depend on the angular velocities separately rather
than on their difference. This question will be studied in the next
section.

%\newpage

\vspace{5mm}

\noindent{\large\bf The Rotating Dielectric Magnet}\hfill

\vspace{3mm}

Assume now that our magnet does not conduct electricity ($\gamma = 0$)
but, instead, electrically polarizable
($\epsilon\ge\epsilon_0$). Then, under stationary conditions, the l.h.s.
of \reff{B1a} is obviously equal to zero but since now $\gamma =0$ Eq.
\reff{B2a} does not follow from this fact.

The field $\boldsymbol E + (\boldsymbol V\times\boldsymbol B)$ which in
a moving conductor determines the current through the Ohm's law makes a 
dielectric polarized:
\begin{equation}
\boldsymbol P = \chi\epsilon_0\bigl [\boldsymbol E + (\boldsymbol V\times\boldsymbol B)\bigr ] =
\boldsymbol P^{(1)} + \boldsymbol P^{(2)}\labl{E1a}
\end{equation}
($\chi$ is the dielectric susceptibility). In the above equation 
$\boldsymbol
P^{(1)} = \chi\epsilon_0\boldsymbol E$ is the electrostatic
while $\boldsymbol P^{(2)} = \chi\epsilon_0(\boldsymbol V\times\boldsymbol B)$
is the magnetically induced (or motional) polarization.

This is, however, not yet the full polarization since there is a third
contribution
\begin{equation}
\boldsymbol P^{(3)} = \epsilon_0\mu_0(\boldsymbol V\times\boldsymbol M)\labl{E2a}
\end{equation}
predicted by relativity theory, according to which elementary
magnetic dipoles $\boldsymbol m$ of a moving permanent magnet acquire
electric dipole moment equal to
$\epsilon_0\mu_0(\boldsymbol V\times\boldsymbol m)$.

The origin of this phenomenon may be understood directly from the
equivalence of the inertial frames of reference without resort to the
apparatus of relativity theory. 

Consider an elementary magnetic dipole $\boldsymbol m$ which is at rest
in an inertial frame of reference and an elementary linear conductor
$d\boldsymbol l$ which is moving with constant velocity $\boldsymbol v$.
The electromotive force $d\cal E$ induced in this elementary conductor
by its motion is equal to
$d{\cal E} = (\boldsymbol v\times\boldsymbol B)\cdot d\boldsymbol l$ in
which $\boldsymbol B$ is the field of the dipole at the position of the
conductor. The absence of any absolute frame of reference requires that
when the conductor is at rest and the dipole is moving with constant
velocity $-\boldsymbol v$ the electromotive force induced remain the
same as before.

The magnetic field of the moving dipole at the fixed position of
$d\boldsymbol l$ varies in time and, therefore, it brings about, through
Maxwell-equations, an electric field $\boldsymbol E$ which in turn
produces an electromotive force $\boldsymbol E\cdot d\boldsymbol l$.
Simple calculation shows that, contrary to the expectation, 
$\boldsymbol E\cdot d\boldsymbol l\not= d{\cal E}$. Equality is obtained
only if in $\boldsymbol E$ one takes into account the electric field of
the electric dipole moment 
$\epsilon_0\mu_0\bigl ((-\boldsymbol v)\times\boldsymbol m)$
acquired by $\boldsymbol m$ due to its velocity $-\boldsymbol v$.

The volume charge densities produced by all three types of polarization
are given by the equation
\begin{equation}
\rho_i = -\Div\boldsymbol P^{(i)}\qquad (i=1,2,3).\labl{E3c}
\end{equation}
Through Gauss-theorem this formula determines the surface charge
densities as
\begin{equation}
\sigma_i = P_r^{(i)}\qquad (r=a,\quad i=1,2,3).\labl{E3d}
\end{equation}
In words: The surface charge densities are given by the normal component of
the polarization vectors on the inner side of the surface.

According to the first Maxwell-equation
\[\Div\epsilon_0\boldsymbol E = \rho_1 + \rho_2 + \rho_3.\]
If we introduce the induction vector by the formula
\[\boldsymbol D = \epsilon_0\boldsymbol E + \boldsymbol P^{(1)}\]
the above equation takes on the form
\begin{equation}
\Div\boldsymbol D = \rho_2 + \rho_3\labl{E4a}
\end{equation}
and we arrive at a standard electrostatic problem: Consider a sphere of
constant dielectric permeability $\epsilon$. Calculate the electrostatic
potential for given volume and surface charge densities
$(\rho_2 + \rho_3)$ and $(\sigma_2 + \sigma_3)$.

Eq. \reff{B3b} and the relation 
$\chi\epsilon_0 = (\epsilon - \epsilon_0)$ permit us to write
\[\boldsymbol P^{(2)} + \boldsymbol P^{(3)} = \frac{1}{3}(2\epsilon + \epsilon_0)(\boldsymbol
V\times\mu_0\boldsymbol M)\]
which in turn leads through \reff{B3c} to
\begin{equation}
\begin{aligned}
\rho_2 + \rho_3 &= -\Div (\boldsymbol P^{(2)} + \boldsymbol P^{(3)}) =\\
&= -\frac{2}{3}(2\epsilon + \epsilon_0)\mu_0M\Omega .
\end{aligned}\labl{E4b}
\end{equation}
This expression will be substituted into the r.h.s. of \reff{E4a}.

Similarly, we obtain for the surface density the equation
\[\sigma_2 + \sigma_3 = P_r^{(2)} + P_r^{(3)} =
\frac{1}{3}(2\epsilon + \epsilon_0)(\boldsymbol V\times\mu_0\boldsymbol M)_r.\]
Using \reff{B3d}, the relation $(\boldsymbol r\cdot\boldsymbol M) =
rM\cos\vartheta$ and $\Omega_r = \Omega\cos\vartheta$ we have
\begin{equation}
\sigma_2 + \sigma_3 = \frac{1}{3}(2\epsilon +
\epsilon_0)a\mu_0M\Omega\sin^2\vartheta .\labl{E5a}
\end{equation}

We write in \reff{E4a} $\boldsymbol D = \epsilon\boldsymbol E$
and through $E = -\boldsymbol\nabla \Phi$ introduce the potential $\Phi$
again. Then
\begin{equation}
\triangle\Phi =
	\begin{cases}
		\di -\frac{1}{\epsilon}(\rho_2+\rho_3)&
		\text{$(r<a)$}\\[2mm]
		0& \text{$(r>a)$.}
	\end{cases}\labl{E5b}
\end{equation}
The boundary condition for $\boldsymbol D$ is fixed by the surface
densities as
\[D_r^+ - D_r^- = \sigma_2 + \sigma_3,\]
in which $\boldsymbol D^+$ and $\boldsymbol D^-$ are the inductions on
the outer and inner sides of the surface. This condition expressed
through the potential becomes
\begin{equation}
-\epsilon_0\frac{\partial\Phi_+}{\partial r} +
\epsilon\frac{\partial\Phi_-}{\partial r} = \sigma_2 + \sigma_3\qquad
(r=a)\labl{E5c}
\end{equation}
($\Phi_+$ and $\Phi_-$ are the potentials outside and inside the sphere).

Eq. \reff{E5b} will be solved in Appendix 3 with the result
\begin{equation}
\begin{aligned}
\Phi = &-\frac{2\epsilon + \epsilon_0}{9(2\epsilon +
3\epsilon_0)}\mu_0M\Omega\: (3\cos^2\vartheta - 1)\times\\[2mm]
&\times
	\begin{cases}
		r^2& \text{$(r<a)$}\\[2mm]
		\di\frac{a^5}{r^3}& \text{$(r>a)$,}
	\end{cases}
\end{aligned}\labl{E6b}
\end{equation}
From this we obtain for the electric part of the electromotive force
$\cal E$ the expression
\begin{equation}
{\cal E}_e = \Phi_B - \Phi_A = \frac{2\epsilon + \epsilon_0}{3(2\epsilon +
3\epsilon_0)}a^2\mu_0M\Omega .\labl{E6c}
\end{equation}

The magnetic part is still given by \reff{D1b}. Hence, the full
electromotive force is given as
\begin{equation}
{\cal E} = \frac{1}{3}a^2\mu_0M\left (\frac{2\epsilon +
\epsilon_0}{2\epsilon + 3\epsilon_0}\Omega - \omega\right ).\labl{E6d}
\end{equation}
As we have already mentioned in this case it is {\em not} 
the relative rotation which determines $\cal E$ .
In spite of the existence of this electromotive force no stationary
current will flow through $\cal L$ since the magnet's conductivity is
zero but the potential difference between the points $A$ and $B$
may be observed.

Since conductors are "infinitely easily" polarizable substances 
\reff{E6d} must be reduced in the
limit $\epsilon\longrightarrow\infty$ to
\reff{D1c} which is indeed the case. 

The numerator of
\reff{E6c} can be written in the form 
$\bigl[2(\epsilon -
\epsilon_0) + 3\epsilon_0\bigr ]$
in which the term $3\epsilon_0$ derives from $\boldsymbol P^{(3)}$.
Since this term does not contain $\epsilon$ it drops out of the limit
$\epsilon\longrightarrow\infty$.
Does this mean that the elementary magnets in a conducting permanent
magnet do not acquire electric dipole moment due to their motion?

Of course, not. According to \reff{B2a}, in a conducting magnet the
rotation determines the electric field directly, independently of
whether the latter is produced by polarization charges or motion-induced
electric dipole moments. In a dielectric magnet, on the contrary, it is
the polarization rather than the field itself which is fixed by the
rotation and so in this case it is crucial to take into account all
possible types of polarizations.

When the magnet is neither conducting ($\gamma =0$) nor polarizable
($\epsilon = \epsilon_0$) the electromotive force in $\cal L$ is still
different from zero:
\[{\cal E} = \frac{1}{3}A^2\mu_0M\left (\frac{3}{5}\Omega - \omega\right
).\]
This electromotive force originates solely from $\boldsymbol P^{(3)}$.

\vspace{5mm}

\noindent{\large\bf The Tethered Satellite}\hfill

\vspace{3mm}

\renewcommand{\thefootnote}{2}

In march of 1996 a spherical 1.6-meter diameter satellite was
released out into space from the payload bay of Space Shuttle Columbia
during its orbiting at a height about 90 km above the Earth. Its tether,
a long conducting cable, served (among others) to generate electric
power due to the electromotive force ${\cal E}_m$ induced in it by the
Erth's magnetic field. Free electrons in the thin ionosphere where the
Space Shuttle operated were attracted to the satellite. The electrons
travelled along the tether to the orbiter. The electric circuit was
closed by means of an electron generator on the orbiter which returned
charged particles back into the ionosphere.

The electromotive force ${\cal E}_m$ is the greater the longer the
tether is. The latter was a 21-kilometer-long leash but it broke when
the satellite was extended 19.7 kilometers. The experiment, however,
could not be considered as a failure. Up to the time of the severing of the
tether, the orbiter-tether-satellite system had been generating 3,500
volts and up to 0.5 amps of current.

Considerations of the preceding sections suggest that, since the Earth
rotates in its own magnetic field, it must be electrically polarized
and the electric field of the polarization charges must give rise to an
electromotive force ${\cal E}_e$ in the tether which also contributes to
the current in it. To have an order of magnitude estimate we assume that
(1) the Earth core is a homogeneous permanent magnet of nonzero
conductivity $\gamma$, (2) the rotational and magnetic axes of the Earth
coincide and (3) the mantle's and the atmosphere's 
polarizabilities are negligible. Neither of these assumptions is correct
but, perhaps, they provide an acceptable starting point.

Consider a tethered orbiter-satellite system, orbiting above the Equator
($xy$ plane) on a circular orbit of radius $r$. Assume further that the
tether, a linear conductor of length $\Delta l\ll r$ is oriented along
the radius and the positive direction on it points toward the increase
of $r$. Then
\[\Delta {\cal E}_m = (\boldsymbol v\times\boldsymbol B)_r\:\Delta l.\]
$\boldsymbol v$ is the orbiting velocity with respect to the
(practically inertial) frame of reference with its origin in the center
of the Earth and orientation defined by the fixed starts. The magnitude of
$\boldsymbol v$ is equal to $\omega r$ where $\omega$ is the angular
velocity of the orbiter. In the orbital plane the Earth's magnetic field
$\boldsymbol B$ has only $z$-component equal to
\[+\frac{1}{3}\mu_0M\left (\frac{a_c}{r}\right )^3\]
where $a_c$ is the radius of the Earth core. Hence 
\[\Delta {\cal E}_m = +\frac{1}{3}\mu_0M\omega\frac{a_c^3}{r^2}\:\Delta
l.\]
This positive $\Delta {\cal E}_m$ 
gives rise to an electron current directed toward the Earth and,
therefore, the satellite has to be orbiting {\em above} the shuttle.

On the other hand,
\[\Delta {\cal E}_e = E_r\:\Delta l = -\frac{\partial\Phi}{\partial
r}\:\Delta l.\]
The r.h.s. is to be calculated at $\vartheta = 90^\circ$.
According to \reff{B5d}
\[\Delta {\cal E}_e = \frac{1}{3}\mu_0M\Omega\frac{a_c^5}{r^4}\:\Delta
l,\]
therefore, 
\[\left\vert\frac{\Delta {\cal E}_e}{\Delta {\cal E}_m}\right\vert = 
\left (\frac{a_c}{r}\right )^2\frac{\Omega}{\omega}\]
where $\Omega$ is the angular velocity of the Earth's rotation. Since
$a_c/r\approx 1/2$ and $\Omega /\omega\approx 1/20$, $\Delta {\cal
E}_e$ is less than $\Delta {\cal
E}_m$ only by about two orders of magnitude.\footnote{
Since charged particles move much faster
than the tethered system,
the effect of the Coulomb-force due to the Earth's polarization
for them is negligible with respect to the Lorentz-force.}
Moreover, if the orbital plane is perpendicular to the Equator (i.e. it
goes through the Poles and is, of course, at rest with respect to
the fixed stars) then, since $\boldsymbol v\times\boldsymbol B$ is perpendicular to 
the radial direction, $\Delta {\cal E}_m = 0$ and it is $\Delta {\cal
E}_e$ {\em alone} which contributes to the electric current in the
tether. Though polarizability of the ionospher may substantially alter
(or even invalidate) this conclusion the possibility of the electric
polarization of the Earth seems worth of further consideration.

\vspace{5mm}

\noindent{\large\bf Appendix 1}\hfill

\vspace{3mm}

We are looking for the solution of the equation \reff{B5c} at $a>r$
which tends to zero faster than $1/r$ (the total charge of the sphere is
zero) and the tangential component of the electric field on the outer
side of the surface $r=a$ coincides with the tangential component of the
field \reff{B2a} inside the sphere. Owing to the surface charges which
compensate the volume charge the normal component of $\boldsymbol E$
will be discontinuous at $r=a$.

Axial symmetry requires $E_\varphi$ to vanish and so the tangential
component is given by
\[E_\vartheta =
-\frac{1}{a}\left.\frac{\partial\Phi}{\partial\vartheta}\right\vert_a\]
alone. According to \reff{B2a} it is equal to
\[E_\vartheta = -(\boldsymbol V\times\boldsymbol B)_\vartheta = -V_\varphi B_r + V_r
B_\varphi .\]
Both factors of the  second term are equal to zero while in the first term
\[V_\varphi = \Omega r\sin\vartheta ,\qquad B_r =
\frac{2}{3}\mu_0M\cos\vartheta .\]
Hence
\begin{equation}
\frac{1}{a}\left.\frac{\partial\Phi}{\partial\vartheta}\right\vert_a =
\frac{2}{3}\mu_0M\Omega a\sin\vartheta\cos\vartheta .\labl{Q1a}
\end{equation}
The axisymmetric solution of \reff{B5c} which obeys all the requirements
formulated is given by the equation
\begin{equation}
\Phi (r,\vartheta ) =\frac{A}{r^3}P_2(\cos\vartheta ) =
\frac{A}{2r^3}(3\cos^2\vartheta - 1)\qquad (r\ge a).\labl{Q2a}
\end{equation}
$P_2(\cos\vartheta )$ is the 2-nd Legendre-polynomial, $A$ is a constant
which is fixed by \reff{Q1a} as
\[A = -\frac{2}{9}a^5\mu_0M\Omega\]
Substituting this into \reff{Q2a} we obtain \reff{B5d}.

\vspace{5mm}

\noindent{\large\bf Appendix 2}\hfill

\vspace{3mm}

In the textbook formula
\[\Rot (\boldsymbol V\times\boldsymbol B) = (\boldsymbol B\cdot\boldsymbol\nabla )\boldsymbol V - (\boldsymbol
V\cdot\boldsymbol\nabla )\boldsymbol B + \boldsymbol V\Div\boldsymbol B - \boldsymbol B\Div\boldsymbol V\]
the last two terms vanish as a consequence of
$\Div\boldsymbol B=0$ and $\Div\boldsymbol V = 0$.
In Cartesian-coordinates 
$\boldsymbol\nabla = (\partial_x,\partial_y,\partial_z)$ and we have
\begin{equation}
\Rot_j(\boldsymbol V\times\boldsymbol B) = B_i\partial_iV_j -
V_i\partial_iB_j\labl{Q3a}
\end{equation}
where summation convention over repeated indices is understood
(e.g. $B_i\partial_i = B_x\partial_x +
B_y\partial_y + B_z\partial_z$).

\renewcommand{\thefootnote}{3}

Consider now Eq. \reff{B1b} . For the components in Cartesian
coordinates we have the formula
$(\boldsymbol\Omega\times\boldsymbol r)_i = \epsilon_{ijk}\Omega_jx_k$
in which $(x_1,x_2,x_3)\equiv (x,y,z)$, and
$\epsilon_{ijk}$ is the fully antisymmetric unit tensor 
($\epsilon$-symbol)\footnote{$\epsilon_{ijk}=+1\;\text{or} -1$
depending on whether $ijk$ is an even or odd permutation of $123$. If
some of the indices $ijk$ coincide the value of the symbol is zero.}.
Using this in \reff{Q3a}, we have
\[\Rot_j(\boldsymbol V\times\boldsymbol B) = \epsilon_{jkl}B_i\partial_i(\Omega_kx_l)
- \epsilon_{ikl}\Omega_kx_l\partial_iB_j.\]
Since the current desity is zero throughout, $\boldsymbol B$ is
rotation-free and $\partial_iB_j = \partial_jB_i$. Hence
\[\Rot_j(\boldsymbol V\times\boldsymbol B) = \epsilon_{jkl}B_i\partial_i(\Omega_kx_l) -
\epsilon_{ikl}\Omega_kx_l\partial_jB_i.\]
In the first term
\[\di\partial_i(\Omega_kx_l) = \Omega_k\frac{\partial x_l}{\partial
x_i} = \Omega\delta_{il}\]
which is equal to $\Omega_k$ at $i=l$ and vanishes at $i\not=l$,
therefore
\[\Rot_j(\boldsymbol V\times\boldsymbol B) = \epsilon_{jki}B_i\Omega_k -
\epsilon_{ikl}\Omega_kx_l\partial_jB_i.\]

In the second term we perform a transformation of the opposite
sense: $x_l\partial_jB_i =
\partial_j(x_lB_i) - \delta_{lj}B_i$. 
If we substitute this into the previous formula and use the constancy of
$\Omega_k$ and the $\epsilon$-symbol we obtain
\[\Rot_j(\boldsymbol V\times\boldsymbol B) = -\partial_j(\epsilon_{ikl}B_i\Omega_kx_l)
+ (\epsilon_{jki} + \epsilon_{ikj})B_i\Omega_k.\]
The sum of the $\epsilon$-symbols is zero and under the sign of partial
derivation we recognize the mixed product of the vectors
$\boldsymbol B,
\boldsymbol\Omega$ and $\boldsymbol r$:
\[\Rot_j(\boldsymbol V\times\boldsymbol B) =
-\partial_j\bigl [\boldsymbol B\cdot (\boldsymbol\Omega\times\boldsymbol r)\bigr ].\]
The mixed product of three vectors is equal to the determinant formed
from their Cartesian-components. Since the magnet is assumed
axisymmetric the vector $\boldsymbol B$ at the point $\boldsymbol r$
lies in the plane defined by $\Omega$ (i.e. the $z$-axis) and the
direction of $\boldsymbol r$. Hence the determinant vanishes and the
field $(\boldsymbol V\times\boldsymbol B)$ is indeed rotationless.
Our proof of this fact is obviously valid for any axisymmetric magnet
which rotates around its symmetry axis.

\vspace{5mm}

\noindent{\large\bf Appendix 3}\hfill

\vspace{3mm}

We are seeking the solution of \reff{E5c} which is everywhere finite and
continuous (even at $r=a$). \reff{E5a} suggests the expected $\vartheta$
dependence of the solution:
\begin{align*}
\Phi_- & = Ar^2P_2(\cos\vartheta ) + f(r) =
\frac{A}{2}r^2(3\cos^2\vartheta - 1) + f(r)\\
\Phi_+ & = \frac{B}{r^3}P_2(\cos\vartheta ) =
\frac{B}{2r^3}(3\cos^2\vartheta - 1)
\end{align*}
in which $A$ and $B$ are constants and $f(r)$ is a particular solution
of
\begin{equation}
\triangle f = \frac{1}{r^2}\frac{d}{dr}\left (r^2\frac{df}{dr}\right ) =
-\frac{1}{\epsilon}(\rho_2 + \rho_3)\labl{Q5a}
\end{equation}
which will be chosen to vanish at $r=a$:
\[f(r) = \frac{\rho_2 + \rho_3}{6\epsilon}(a^2 - r^2).\]
Continuity requires $B=Aa^5$, hence
\begin{align*}
\Phi_- & = \frac{A}{2}r^2(3\cos^2\vartheta - 1) + \frac{\rho_2 +
\rho_3}{6\epsilon}(a^2 - r^2)\\
\Phi_+ & = \frac{a^5A}{2r^3}(3\cos^2\vartheta - 1).
\end{align*}
Using \reff{E5a}, the r.h.s. of \reff{E5c} may be written in the form
\[\sigma_2 + \sigma_3 = \frac{1}{9}(2\epsilon + \epsilon_0)a\mu_0M\Omega
[2 - (3\cos^2\vartheta - 1)].\]
Then \reff{E5c} becomes
\begin{equation}
\begin{gathered}
\epsilon_0\cdot\frac{3}{2}aA(3\cos^2\vartheta - 1) +
\epsilon aA(3\cos^2\vartheta - 1) - \frac{1}{3}(\rho_2 + \rho_3)a =\\
= \frac{1}{9}(2\epsilon + \epsilon_0)a\mu_0M\Omega[2 - (3\cos^2\vartheta -
1)]
\end{gathered}\labl{Q5b}
\end{equation}
As it follows from \reff{E4b} the
terms which do not contain $(3\cos^2\vartheta -1)$ cancel
and the remaining equation determines the value of $A$:
\[A = -\frac{2}{9}\cdot\frac{2\epsilon + \epsilon_0}{2\epsilon +
3\epsilon_0}\mu_0M\Omega .\]
From this \reff{E6b} follows.
The cancellation of the terms which do not contain $P_2(\cos\vartheta )$ is the
consequence of the fact that, according to \reff{E3c} and \reff{E3d},
the volume and surface charges compensate to zero separately for all
three types of polarization charges.

\end{document}